\documentclass[12pt]{article}
\usepackage{amssymb,amscd,array}
\catcode `\@=11
\@addtoreset{equation}{section}

\newtheorem{thm}{Theorem}[section]

\newtheorem{lemma}[thm]{Lemma}
\newtheorem{prop}[thm]{Proposition}
\newtheorem{cor}[thm]{Corollary}

\def\qed{\blacksquare}
\newcommand{\be}{\begin{equation}}
\newcommand{\ee}{\end{equation}}
\newcommand{\bea}{\begin{eqnarray}}
\newcommand{\eea}{\end{eqnarray}}
\newcommand{\R}{\mathbb{R}}
\newcommand{\N}{\mathbb{N}}
\newcommand{\C}{\mathbb{C}}

\begin{document}
\begin{titlepage}

\begin{center}
{\bf \Large{A Generalization of Gauge Invariance\\}}
\end{center}
\vskip 1.0truecm
\centerline{D. R. Grigore, 
\footnote{e-mail: grigore@theory.nipne.ro}}
\vskip5mm
\centerline{Department of Theoretical Physics, Institute for Physics and Nuclear
Engineering ``Horia Hulubei"}
\centerline{Institute of Atomic Physics}
\centerline{Bucharest-M\u agurele, P. O. Box MG 6, ROM\^ANIA}

\vskip 2cm
\bigskip \nopagebreak
\begin{abstract}
\noindent
We consider perturbative quantum field theory in the causal framework. Gauge invariance is, in this framework, an identity
involving chronological products of the interaction Lagrangian; it express the fact that the scattering matrix must
leave invariant the sub-space of physical states. We are interested in generalizations of such identity involving Wick sub-monomials 
of the interaction Lagrangian. The analysis can be performed by direct computation in the lower orders of perturbation theory; guided 
by these computations we conjecture a generalization for arbitrary orders.
\end{abstract}
\end{titlepage}

\section{Introduction}

The general framework of perturbation theory consists in the construction of 
the chronological products such that Bogoliubov axioms are verified 
\cite{BS}, \cite{EG}, \cite{DF}; for every set of Wick monomials 
$ 
A_{1}(x_{1}),\dots,A_{n}(x_{n}) 
$
acting in some Fock space
$
{\cal H}
$
one associates the operator
$$ 
T(A_{1}(x_{1}),\dots,A_{n}(x_{n})) 
$$ 
which is a distribution-valued operators called chronological product. 

The construction of the chronological products can be done recursively according
to Epstein-Glaser prescription \cite{EG}, \cite{Gl} (which reduces the induction
procedure to a distribution splitting of some distributions with causal support)
or according to Stora prescription \cite{PS} (which reduces the renormalization
procedure to the process of extension of distributions). These products are not
uniquely defined but there are some natural limitation on the arbitrariness. If
the arbitrariness does not grow with $n$ we have a renormalizable theory. A variant
based on retarded products is due to Steinmann \cite{Sto1}.

Gauge theories describe particles of higher spin. Usually such theories are not
renormalizable. However, one can save renormalizablility using ghost fields.
Such theories are defined in a Fock space
$
{\cal H}
$
with indefinite metric, generated by physical and un-physical fields (called
{\it ghost fields}). One selects the physical states assuming the existence of
an operator $Q$ called {\it gauge charge} which verifies
$
Q^{2} = 0
$
and such that the {\it physical Hilbert space} is by definition
$
{\cal H}_{\rm phys} \equiv Ker(Q)/Im(Q).
$
The space
$
{\cal H}
$
is endowed with a grading (usually called {\it ghost number}) and by
construction the gauge charge is raising the ghost number of a state. Moreover,
the space of Wick monomials in
$
{\cal H}
$
is also endowed with a grading which follows by assigning a ghost number to
every one of the free fields generating
$
{\cal H}.
$
The graded commutator
$
d_{Q}
$
of the gauge charge with any operator $A$ of fixed ghost number
\be
d_{Q}A = [Q,A]
\ee
is raising the ghost number by a unit. It means that
$
d_{Q}
$
is a co-chain operator in the space of Wick polynomials. From now on
$
[\cdot,\cdot]
$
denotes the graded commutator. From
$
Q^{2} = 0
$
one derives
\be
(d_{Q})^{2} = 0.
\label{Q-square}
\ee
 
A gauge theory assumes also that there exists a Wick polynomial of null ghost
number
$
T(x)
$
called {\it the interaction Lagrangian} such that
\be
d_{Q}T = [Q, T] = i \partial_{\mu}T^{\mu}
\label{gauge-T}
\ee
for some other Wick polynomials
$
T^{\mu}.
$
This relation means that the expression $T$ leaves invariant the physical
states, at least in the adiabatic limit. Indeed, if this is true we have:
\be
T(f)~{\cal H}_{\rm phys}~\subset~~{\cal H}_{\rm phys}  
\label{phys-inv}
\ee
up to terms which can be made as small as desired (making the test function $f$
flatter and flatter). We call this argument the {\it formal adiabatic limit}.
It is a way to justify from the physical point of view relation (\ref{gauge-T}). Otherwise,
we simply have to postulate it. The preceding relation can be extended if we assume a polynomial
Poincar\'e lemma as follows. One applies 
$
d_{Q}
$
to (\ref{gauge-T}) and obtains
\bea
\partial_{\mu}d_{Q}T^{\mu} = 0
\nonumber
\eea
so we expect that we have
\bea
d_{Q}T^{\mu} = i \partial_{\nu}T^{\mu\nu}
\eea
and so on. 

One defines now the chronological products
$
T(A_{1}(x_{1}),\dots,A_{n}(x_{n})) 
$
with 
$
A_{1},\dots,A_{n}
$
of the type
$
T, T^{\mu}, T^{\mu\nu},
$
etc. and formulates a proper generalization of (\ref{gauge-T}). Such identity express, as (\ref{phys-inv}), the fact that
the scattering matrix leaves invariant the subspace of physical states, at least in some adiabatic limit sense. The analysis
of these identities can be done by direct computations in lower orders of the perturbation theory, but a general proof in
arbitrary orders is still an open problem in the general case, due to the quantum anomalies which do appear in the inductive
procedure.

The purpose of this paper is to investigate the uses of the so-called Wick submonomials. These expressions appear for the 
first time in the original paper of Epstein and Glaser and they are used to express normal and chronological products
in terms of numerical distributions multiplying Wick products - see formula (41) and (42) from \cite{EG}. A more elaborate
way to use Wick submonomials appears in \cite{DF}. We will use in the following a variant of \cite{DF} approach appearing 
in \cite{Bo}, \cite{Bo+Du} and we want to
investigate the possibility to express gauge invariance (\ref{s-gauge-n}) in terms of numerical distributions. We analyze 
the lowest orders of perturbation theory ($n = 2$) and obtain such relations. We provide a conjecture for the higher orders
of perturbation theory and a new strategy to analyze anomalies in the general case. Another variant of this analysis was 
proposed in \cite{ward}.

In the next two Sections we provide our version of the construction of Wick products and of Bogoliubov axioms. 
Then we consider Wick submonomials of the simplest gauge theory namely of QCD type in Sections \ref{ym} and \ref{ws}. Finally,
in Section \ref{rank} we give our conjecture and a strategy to derive gauge invariance from ``simpler" identities involving
Wick submonomials of
$
T, T^{\mu}, T^{\mu\nu},
$
etc.

\newpage

\section{Wick Products\label{wick prod}}
We consider a classical field theory on the Minkowski space
$
{\cal M} \simeq \R^{4}
$
(with variables
$
x^{\mu}, \mu = 0,\dots,4)
$
described by the Grassmann manifold 
$
\Xi_{0}
$
with variables
$
\xi_{a}, a \in {\cal A}
$
(here ${\cal A}$ is some index set) and the associated jet extension
$
J^{r}({\cal M}, \Xi_{0}),~r \geq 1
$
with variables 
$
x^{\mu},~\xi_{a;\mu_{1},\dots,\mu_{n}},~n = 0,\dots,r;
$
we denote generically by
$
\xi_{p}, p \in P
$
the variables corresponding to classical fields and their formal derivatives. 
These variables generate the algebra
$
\Xi
$
of polynomials.

To illustrate this, let us consider a real scalar field in Minkowski space ${\cal M}$. The first jet-bundle extension is
$$
J^{1}({\cal M}, \R) \simeq {\cal M} \times \R \times \R^{4}
$$
with coordinates 
$
(x^{\mu}, \phi, \phi_{\mu}),~\mu = 0,\dots,3.
$

If 
$
\varphi: \cal M \rightarrow \R
$
is a smooth function we can associate a new smooth function
$
j^{1}\varphi: {\cal M} \rightarrow J^{1}(\cal M, \R) 
$
according to 
$
j^{1}\varphi(x) = (x^{\mu}, \varphi(x), \partial_{\mu}\varphi(x)).
$

For higher order jet-bundle extensions we have to add new real variables
$
\phi_{\{\mu_{1},\dots,\mu_{r}\}}
$
considered completely symmetric in the indexes. For more complicated fields, one needs to add supplementary indexes to
the field i.e.
$
\phi \rightarrow \phi_{a}
$
and similarly for the derivatives. The index $a$ carries some finite dimensional representation of
$
SL(2,\C)
$
(Poincar\'e invariance) and, maybe a representation of other symmetry groups. 
In classical field theory the jet-bundle extensions
$
j^{r}\varphi(x)
$
do verify Euler-Lagrange equations. To write them we need the formal derivatives defined by
\be
\partial_{\nu}\phi_{\{\mu_{1},\dots,\mu_{r}\}} \equiv \phi_{\{\nu,\mu_{1},\dots,\mu_{r}\}}.
\ee

We suppose that in the algebra 
$
\Xi
$
generated by the variables 
$
\xi_{p}
$
there is a natural conjugation
$
A \rightarrow A^{\dagger}.
$
If $A$ is some monomial in these variables, there is a canonical way to associate to $A$ a Wick 
monomial: we associate to every classical field
$
\xi_{a}, a \in {\cal A}
$
a quantum free field denoted by
$
\xi^{\rm quant}_{a}(x), a \in {\cal A}
$
and determined by canonical (anti)commutation relations:
\be
[ \xi^{\rm quant}_{a}(x), \xi^{\rm quant}_{b}(y) ] = 
- i~D( \xi_{a}(x), \xi_{b}(y) )\times {\bf 1} = - i~D_{ab}(x - y )\times {\bf 1}
\label{CCR}
\ee
where 
$
[\cdot,\cdot ]
$
is the graded commutator and 
$
D_{ab}(x)
$
is the causal Pauli-Jordan distribution associated to the two quantum fields; it is (up to some numerical factors) a polynomial
in the derivatives applied to the Pauli-Jordan distribution. Afterwards we define
$$
\xi^{\rm quant}_{a;\mu_{1},\dots,\mu_{n}}(x) \equiv \partial_{\mu_{1}}\dots \partial_{\mu_{n}}\xi^{\rm quant}_{a}(x), a \in {\cal A}
$$
and extend (\ref{CCR}) to all fields
$
\xi^{\rm quant}_{p}
$
associated to
$
\xi_{p} \in \Xi.
$
To compute the graded commutator one needs the Fermi
parities
$
|\xi_{p}|
$
of the fields
$
\xi_{p}, p \in P.
$
The distributions
$
D_{pq}(x)
$
can be split into a positive (resp. negative) frequency part
$
D_{pq}^{(\pm)}(x).
$
The free quantum fields are generating a Fock space 
$
{\cal F}
$
in the sense of the Borchers algebra: formally it is generated by states of the form
$
\xi^{\rm quant}_{a_{1}}(x_{1})\dots \xi^{\rm quant}_{a_{n}}(x_{n})\Omega
$
where 
$
\Omega
$
the vacuum state.
The scalar product in this Fock space is constructed using the $2$-point distributions
$
D^{(+)}_{pq}(x)
$
and we denote by
$
{\cal F}_{0} \subset {\cal F}
$
the algebraic Fock space.

 Because the quantum fields are supposed free, this means that 
they verify some free field equation; in particular every field must verify Klein Gordon equation for some mass $m$
\be
(\square + m^{2})~\xi^{\rm quant}_{a}(x) = 0
\ee
and it follows that in momentum space they must have the support on the hyperboloid of mass $m$. This means that 
they can be split in two parts
$
\xi^{\rm quant (\pm)}_{a}
$
with support on the upper (resp.lower) hyperboloid of mass $m$. We convene that 
$
\xi^{\rm quant (+)}_{a} 
$
resp.
$
\xi^{\rm quant (-)}_{a} 
$
correspond to the creation (resp. annihilation) part of the quantum field. The expressions
$
\xi^{\rm quant (+)}_{p} 
$
resp.
$
\xi^{\rm quant (-)}_{p} 
$
for a generic
$
\xi_{p},~ p \in P
$
are obained in a natural way, applying partial derivatives.

Then we associate to a monomial $A$ in the variables from
$
\Xi
$
the Wick monomial
$
A^{\rm quant}(x)
$
by replacing
$$
\xi_{a;\mu_{1},\dots,\mu_{n}} \rightarrow \partial_{\mu_{1}}\dots \partial_{\mu_{n}}\xi^{\rm quant}_{a}(x)
$$
(without changing the order of the factors) and then, applying Wick ordering. There is a obvious reverse process
of obtaining from a Wick monomial 
$
A^{\rm quant}(x)
$
the polynomial $A$ depending on variables from
$
\Xi
$
i.e. there is a biunivoc relation
$
A \leftrightarrow A^{\rm quant}(x).
$
If it is obvious from the 
context that we are referring to quantum fields we abandon, for simplicity, the superscript ``quant".
The Wick monomials are leaving invariant the algebraic Fock space.

Let us provide the definition for the Wick monomials.

\begin{prop}
The operator-valued distributions
$
N(\xi_{q_{1}}(x_{1}),\dots,\xi_{q_{n}}(x_{n}))
$
are uniquely defined by:

(i) 
\be
N(\xi_{q_{1}}(x_{1}),\dots,\xi_{q_{n}}(x_{n}))\Omega = \xi_{q_{1}}^{(+)}(x_{1})\dots  \xi_{q_{n}}^{(+)}(x_{n})\Omega
\ee

(ii)
\bea
[ \xi_{p}(y), N(\xi_{q_{1}}(x_{1}),\dots,\xi_{q_{n}}(x_{n})) ] = 
\nonumber\\
- i~\sum_{m=1}^{n} \prod_{l <m} (-1)^{|\xi_{p}||\xi_{q_{l}}|}~D_{pq_{m}}(y - x_{m})~N(\xi_{q_{1}}(x_{1}),\dots,\hat{m},\dots,\xi_{q_{n}}(x_{n}))
\eea

(iii)
\be
N(\emptyset) = I.
\ee

The expression
$
N(\xi_{q_{1}}(x_{1}),\dots,\xi_{q_{n}}(x_{n}))
$
is (graded) symmetrical in the arguments.
\end{prop}

{\bf Proof:} First we use (ii) for $ n = 1$ and (iii) to prove that
\be
N(\xi_{q}(x)) = \xi_{q}(x)
\ee
and then we use the fact that the Fock space is generated by states of the form
$$
\xi_{p_{1}}(x_{1})\dots  \xi_{p_{m}}(x_{m})\Omega.
$$

We can compute the state
$
N(\xi_{q_{1}}(x_{1}),\dots,\xi_{q_{n}}(x_{n}))~\xi_{p_{1}}(x_{1})\dots  \xi_{p_{m}}(x_{m})\Omega
$
by commuting the expression 
$
N(\xi_{q_{1}}(x_{1}),\dots,\xi_{q_{n}}(x_{n}))
$
successively with the $m$ quantum fields and then we use
$
N(\xi_{q_{1}}(x_{1}),\dots,\xi_{q_{n}}(x_{n}))\Omega
$
from (i).

For the last assertion one commutes an arbitrary quantum field with the difference 
$$
N(\xi_{q_{1}}(x_{1}),\dots,\xi_{q_{n}}(x_{n}))
- (-1)^{|\xi_{j}||\xi_{j+1}|}~N(\xi_{q_{1}}(x_{1}),\dots,\xi_{q_{j+1}}(x_{j+1}),\xi_{q_{j}}(x_{j}),\dots,\xi_{q_{n}}(x_{n}))
$$
and gets zero.
$\qed$

The expression
$
N(\xi_{q_{1}}(x_{1}),\dots,\xi_{q_{n}}(x_{n}))
$
are called {\it Wick monomials}. 

Next we have 
\begin{prop}
The following relation is true:
\be
N(\xi_{q_{1}}(x_{1}),\dots,\xi_{q_{n}}(x_{n})) =
\sum_{I,J \in Part(N)}~\epsilon(I,J)~\prod_{i \in I} \xi_{q_{i}}^{(+)}(x_{i})~\prod_{j \in J} \xi_{q_{j}}^{(+)}(x_{j})
\label{N-pm}
\ee
where
$
N = \{1,\dots,n\}
$
and the subsets $I,J$ are ordered. We have defined the sign
\be
\epsilon(I,J) = (-1)^{s}
\ee
where
\be
s \equiv \sum_{i \in I} |\xi_{q_{i}}| \sum_{j \in J,j < i} |\xi_{q_{j}}|
\ee
\end{prop}
{\bf Proof:} We denote the right hand side of the (\ref{N-pm}) by
$
N^{\prime}(\xi_{q_{1}}(x_{1}),\dots,\xi_{q_{n}}(x_{n}))
$
and prove that it verifies (i) - (iii) of the preceding proposition. Only (ii) is highly non-trivial. It is better to 
commute 
$
N^{\prime}(\xi_{q_{1}}(x_{1}),\dots,\xi_{q_{n}}(x_{n}))
$
with
$
\xi^{(+)} 
$
and
$
\xi^{(-)} 
$
and to add the result. 
$\qed$

As a byproduct we get:
\begin{cor}
The following formula is true:
\bea
[ \xi^{(\epsilon)}_{p}(y), N(\xi_{q_{1}}(x_{1}),\dots,\xi_{q_{n}}(x_{n})) ] = 
\nonumber\\
- i~\sum_{m=1}^{n} \prod_{l <m} (-1)^{|\xi_{p}||\xi_{q_{l}}|}~D^{(-\epsilon)}_{pq_{m}}(y - x_{m})~N(\xi_{q_{1}}(x_{1}),\dots,\hat{m},\dots,\xi_{q_{n}}(x_{n}))
\eea
where 
$
\epsilon = \pm
$
and
$
D^{(\pm)}_{pq}
$
are the positive (resp. negative) frequency parts of
$
D^{(\pm)}.
$
\end{cor}

It is a non-trivial result of Wightman and Garding \cite{WG} that in
$
N(\xi_{q_{1}}(x_{1}),\dots,\xi_{q_{n}}(x_{n}))
$
one can collapse all variables into a single one and still gets an well-defined expression: 
\begin{prop}
The expressions
\be
W_{q_{1},\dots,q_{n}}(x) \equiv N(\xi_{q_{1}}(x),\dots,\xi_{q_{n}}(x))
\ee
are well-defined. They verify:
(i) 
\be
W_{q_{1},\dots,q_{n}}(x) \Omega = \xi_{q_{1}}^{(+)}(x)\dots  \xi_{q_{n}}^{(+)}(x)\Omega
\ee

(ii)
\bea
[ \xi^{(\epsilon)}_{p}(y), W_{q_{1},\dots,q_{n}}(x)  ] = 
- i~\sum_{m=1}^{n} \prod_{l <m} (-1)^{|\xi_{p}||\xi_{q_{l}}|}~D_{pq_{m}}^{(-\epsilon)}(y - x_{m})~W_{q_{1},\dots,\hat{m},\dots,q_{n}}(x)
\eea

(iii)
\be
W(\emptyset) = I.
\ee
\end{prop}
We call expressions of the type
$
W_{q_{1},\dots,q_{n}}(x) 
$
{\it Wick monomials}.  By
\be
|W| \equiv \sum_{l=1}^{n} |\xi_{q_{l}}|
\ee
we mean the Fermi number of $W$. We define the derivative
\be
{\partial \over \partial \xi_{p}}W_{q_{1},\dots,q_{n}}(x) \equiv 
\sum_{s=1}^{n}~\prod_{l < s}~(-1)^{|\xi_{p}||\xi_{q_{l}}|}~\delta_{pq_{s}}~W_{q_{1},\dots,\hat{q_{s}},\dots,q_{n}}(x)
\ee
and we have a generalization of the preceding Proposition.
\begin{prop}
Let
$
W_{j} = W_{q^{(j)}_{1},\dots,q^{(j)}_{r_{j}}},~j = 1,\dots,n
$
be Wick monomials. Then the expression
$
N(W_{1}(x_{1}),\dots,W_{n}(x_{n}))
$
is well-defined through
(i) 
\be
N(W_{1}(x_{1}),\dots,W_{n}(x_{n}))\Omega = \prod_{i=1}^{n} \prod_{l=1}^{r_{j}} \xi_{q^{(j)}_{l}}^{(+)}(x_{i})\Omega
\ee

(ii)
\bea
[ \xi_{p}(y), N(W_{1}(x_{1}),\dots,W_{n}(x_{n}))  ] = 
\nonumber\\
- i~\sum_{m=1}^{n} \prod_{l <m} (-1)^{|\xi_{p}||W_{l}|}~\sum_{q}~D_{pq}(y - x_{m})~
N(W_{1}(x_{1}),\dots,{\partial \over \partial \xi_{q}}W_{m}(x_{m}),\dots,W_{n}(x_{n}))
\eea

(iii)
\be
N(W_{1}(x_{1}),\dots,W_{n}(x_{n}),{\bf 1}) = N(W_{1}(x_{1}),\dots,W_{n}(x_{n})) 
\ee

(iv)
\be
N(W(x)) = W(x).
\ee

The expression
$
N(W_{1}(x_{1}),\dots,W_{n}(x_{n}))
$
is symmetric (in the Grassmann sense) in the entries
$
W_{1}(x_{1}),\dots,W_{n}(x_{n})
$
and verifies
\bea
[ \xi_{p}^{(\epsilon)}(y), N(W_{1}(x_{1}),\dots,W_{n}(x_{n}))  ] = 
\nonumber\\
- i~\sum_{m=1}^{n} \prod_{l <m} (-1)^{|\xi_{p}||W_{l}|}~\sum_{q}~D_{pq}^{(-\epsilon)}(y - x_{m})~
N(W_{1}(x_{1}),\dots,{\partial \over \partial \xi_{q}}W_{m}(x_{m}),\dots,W_{n}(x_{n}))
\eea
\end{prop}

Now we are ready for the most general setting. If $A$ a monomial in the variables of the algebra
$
\Xi
$
we define
\be
\xi \cdot A \equiv (-1)^{|\xi| |A|}~{\partial \over \partial \xi}A
\label{derivative}
\ee
for all
$
\xi \in \Xi_{0}.
$
Here 
$|A|$ 
is the Fermi parity of $A$ and we consider the left derivative in the Grassmann sense; the extra sign will be justified later. Then we have:
\begin{thm}
Let 
$
A_{1},\dots,A_{n}
$
be Grassmann monomials in the variables
$
\xi_{p}, \xi_{p;\mu_{1},\dots,\mu_{n}}
$
from
$
\Xi.
$
Then the expressions
$
N(A_{1}(x_{1}),\dots,A_{n}(x_{n}))
$
are well defined through:

(i) 
\be
N(A_{1}(x_{1}),\dots,A_{n}(x_{n}))\Omega = \prod_{l=1}^{n} A_{l}^{(+)}(x_{l})\Omega
\ee
where
$
A_{l}^{(+)} = A_{l}^{\rm quant (+)}
$
is obtained from
$
A_{l}
$
with the substitutions
$
\xi_{a;\mu_{1},\dots,\mu_{n}} \rightarrow \xi^{\rm quant (+)}_{a;\mu_{1},\dots,\mu_{n}} 
$
and preserving the order of the factors.

(ii)
\bea
[ \xi_{p}(y), N(A_{1}(x_{1}),\dots,A_{n}(x_{n}))  ] = 
\nonumber\\
- i~\sum_{m=1}^{n} \prod_{l \leq m} (-1)^{|\xi_{p}||A_{l}|}~\sum_{q}~D_{pq}(y - x_{m})~
N(A_{1}(x_{1}),\dots,\xi_{q}\cdot A_{m}(x_{m}),\dots,A_{n}(x_{n}))
\label{comm-wick}
\eea

(iii)
\be
N(A_{1}(x_{1}),\dots,A_{n}(x_{n}),{\bf 1}) = N(A_{1}(x_{1}),\dots,A_{n}(x_{n})) 
\ee

(iv)
\be
N(A(x)) = A^{\rm quant}(x).
\ee
The expression
$
N(A_{1}(x_{1}),\dots,A_{n}(x_{n}))
$
is symmetric (in the Grassmann sense) in the entries
$
A_{1}(x_{1}),\dots,A_{n}(x_{n})
$
and verifies
\bea
[ \xi_{p}^{(\epsilon)}(y), N(A_{1}(x_{1}),\dots,A_{n}(x_{n}))  ] = 
\nonumber\\
- i~\sum_{m=1}^{n} \prod_{l \leq m} (-1)^{|\xi_{p}||A_{l}|}~\sum_{q}~D_{pq}^{(-\epsilon)}(y - x_{m})~
N(A_{1}(x_{1}),\dots,\xi_{q}\cdot A_{m}(x_{m}),\dots,A_{n}(x_{n}))
\eea
\end{thm}

An expression
$
E(A_{1}(x_{1}),\dots,A_{n}(x_{n}))
$
is called {\it of Wick type} iff verifies:

\bea
[ \xi_{p}(y), E(A_{1}(x_{1}),\dots,A_{n}(x_{n}))  ] = 
\nonumber\\
- i~\sum_{m=1}^{n} \prod_{l \leq m} (-1)^{|\xi_{p}||A_{l}|}~\sum_{q}~D_{pq}(y - x_{m})~
E(A_{1}(x_{1}),\dots,\xi_{q}\cdot A_{m}(x_{m}),\dots,A_{n}(x_{n}))
\eea

\be
E(A_{1}(x_{1}),\dots,A_{n}(x_{n}),{\bf 1}) = E(A_{1}(x_{1}),\dots,A_{n}(x_{n})) 
\ee

\be
E(1) = {\bf 1}.
\ee
Then we easily have:
\begin{prop}
If
$
E(A_{1}(x_{1}),\dots,A_{k}(x_{k}))
$
and
$
F(A_{k+1}(x_{1}),\dots,A_{n}(x_{n}))
$
are expressions of Wick type, then
$
E(A_{1}(x_{1}),\dots,A_{k}(x_{k}))~F(A_{k+1}(x_{1}),\dots,A_{n}(x_{n}))
$
is also an expression of Wick type.
\end{prop}

Now we formulate Wick theorem. First we extend the product (\ref{derivative}) to more factors through
\be
(\xi\eta)\cdot A \equiv \xi\cdot (\eta\cdot A),\quad \xi,\eta \in \Xi_{0}
\ee
and $A$ an arbitrary monomial. In particular it makes sense to consider expressions of the type
$
B\cdot A
$
where $A$ and $B$ are both monomials. One gets something non-null if $B$ is a {\it submonomial} of $A$. One easily derives that 
\be
A\cdot A = C(A) {\bf 1}
\ee
where 
$
C(A)
$
is a numerical factor. Then we have:
\begin{thm} ({\bf Wick})
Let
$
E(A_{1}(x_{1}),\dots,A_{n}(x_{n}))
$
be an expression of Wick type. The following formula is true:
\bea
E(A_{1}(x_{1}),\dots,A_{n}(x_{n})) = 
\sum_{B \in \Xi}~\epsilon(B_{1},\dots,B_{n};A_{1},\dots,A_{n})
\nonumber\\
<\Omega, E(B_{1}(x_{1}),\dots,B_{n}(x_{n}))\Omega>~
N(B_{1}\cdot A_{1}(x_{1}),\dots,B_{n}\cdot A_{n}(x_{n}))
\label{wick-thm}
\eea
where 
$
B_{j}
$
are distinct Wick submonomials of
$
A_{j}
$
and
\be
\epsilon(B_{1},\dots,B_{n};A_{1},\dots,A_{n}) \equiv (-1)^{s}~\prod_{l=1}^{n}~C(B_{l})^{-1}
\ee
with 
\be
s \equiv \sum_{l=1}^{n} |B_{l}|~(\sum_{p=l+1}^{n}~(|A_{p}| + |B_{p}|) = \sum_{p=2}^{n}~(|A_{p}| + |B_{p}|)~(\sum_{l=1}^{p-1}~|B_{l}|). 
\ee
\end{thm}
{\bf Proof:} It is done by induction over
$
d \equiv deg(A_{1}) + \cdots deg(A_{n}).
$
For 
$
d = 1
$
the assertion of the theorem is trivial. We suppose that it is true for 
$
d < r
$
and we prove it for 
$
d = r.
$
We consider that we have 
$
d = r
$
in (\ref{wick-thm}) and commute this expression with an arbitrary
$
\xi_{p}(y).
$
Using the induction hypothesis we get zero, so formula (\ref{wick-thm}) must be true up to a constant term. Taking the vacuum average we get
that the constant is in fact zero.
$\qed$

In the same way we prove:
\begin{thm}
The following formula is true:
\bea
N(\xi_{p}(y),A_{1}(x_{1}),\dots,A_{n}(x_{n})) = \xi_{p}(y)~N(A_{1}(x_{1}),\dots,A_{n}(x_{n}))
\nonumber\\
+ i~\sum_{m=1}^{n} \prod_{l <m} (-1)^{|\xi_{p}||A_{l}|}~\sum_{q}~D_{pq}(y - x_{m})~
N(A_{1}(x_{1}),\dots,\xi_{q}\cdot A_{m}(x_{m}),\dots,A_{n}(x_{n}))
\eea
\end{thm}

\newpage
\section{Bogoliubov Axioms \label{Bogoliubov}}
Suppose the monomials
$
A_{1},\dots,A_{n}
$
are self-adjoint:
$
A_{j}^{\dagger} = A_{j},~\forall j = 1,\dots,n
$
and of Fermi number
$
f_{i}.
$
We impose for the quantum associated Wick monomials the {\it causality} property:
\be
A^{\rm quant}_{j}(x)~A^{\rm quant}_{k}(y) = (- 1)^{f_{j}f_{k}}~~A^{\rm quant}_{k}(y)~A^{\rm quant}_{j}(x)
\ee
for 
$
(x - y)^{2} < 0
$
i.e.
$
x - y
$
outside the causal cones (this relation is denoted by
$
x \sim y
$).

The chronological products
$$ 
T(A_{1}(x_{1}),\dots,A_{n}(x_{n})) \equiv T^{A_{1},\dots,A_{n}}(x_{1},\dots,x_{n}) \quad n = 1,2,\dots
$$
are some distribution-valued operators leaving invariant the algebraic Fock space and verifying the following set of axioms:
\begin{itemize}
\item
{\bf Skew-symmetry} in all arguments:
\be
T(\dots,A_{i}(x_{i}),A_{i+1}(x_{i+1}),\dots,) =
(-1)^{f_{i} f_{i+1}} T(\dots,A_{i+1}(x_{i+1}),A_{i}(x_{i}),\dots)
\ee

\item
{\bf Poincar\'e invariance}: we have a natural action of the Poincar\'e group in the
space of Wick monomials and we impose that for all 
$g \in inSL(2,\C)$
we have:
\be
U_{g} T(A_{1}(x_{1}),\dots,A_{n}(x_{n})) U^{-1}_{g} =
T(g\cdot A_{1}(x_{1}),\dots,g\cdot A_{n}(x_{n}))
\label{invariance}
\ee
where in the right hand side we have the natural action of the Poincar\'e group on
$
\Xi
$.

Sometimes it is possible to supplement this axiom by other invariance
properties: space and/or time inversion, charge conjugation invariance, global
symmetry invariance with respect to some internal symmetry group, supersymmetry,
etc.
\item
{\bf Causality}: if 
$
x - y 
$
is in the upper causal cone then we denote this relation by
$
x \succeq y
$.
Suppose that we have 
$x_{i} \succeq x_{j}, \quad \forall i \leq k, \quad j \geq k+1$.
then we have the factorization property:
\be
T(A_{1}(x_{1}),\dots,A_{n}(x_{n})) =
T(A_{1}(x_{1}),\dots,A_{k}(x_{k}))~~T(A_{k+1}(x_{k+1}),\dots,A_{n}(x_{n}));
\label{causality}
\ee

\item
{\bf Unitarity}: We define the {\it anti-chronological products} using a convenient notation introduced
by Epstein-Glaser, adapted to the Grassmann context. If 
$
X = \{j_{1},\dots,j_{s}\} \subset N \equiv \{1,\dots,n\}
$
is an ordered subset, we define
\be
T(X) \equiv T(A_{j_{1}}(x_{j_{1}}),\dots,A_{j_{s}}(x_{j_{s}})).
\ee
Let us consider some Grassmann variables
$
\theta_{j},
$
of parity
$
f_{j},  j = 1,\dots, n
$
and let us define
\be
\theta_{X} \equiv \theta_{j_{1}} \cdots \theta_{j_{s}}.
\ee
Now let
$
(X_{1},\dots,X_{r})
$
be a partition of
$
N = \{1,\dots,n\}
$
where
$
X_{1},\dots,X_{r}
$
are ordered sets. Then we define the sign
$
\epsilon(X_{1},\dots,X_{r})
$
through the relation
\be
\theta_{1} \cdots \theta_{n} = \epsilon(X_{1}, \dots,X_{r})~\theta_{X_{1}} \dots \theta_{X_{r}}
\ee
and the antichronological products are defined according to
\be
(-1)^{n} \bar{T}(N) \equiv \sum_{r=1}^{n} 
(-1)^{r} \sum_{I_{1},\dots,I_{r} \in Part(N)}
\epsilon(X_{1},\dots,X_{r})~T(X_{1})\dots T(X_{r})
\label{antichrono}
\ee
Then the unitarity axiom is:
\be
\bar{T}(N) = T(N)^{\dagger}.
\label{unitarity}
\ee
\item
{\bf The ``initial condition"}:
\be
T(A(x)) = A^{\rm quant}(x).
\ee

\item
{\bf Power counting}: We can also include in the induction hypothesis a limitation on the order of
singularity of the vacuum averages of the chronological products associated to
arbitrary Wick monomials
$A_{1},\dots,A_{n}$;
explicitly:
\be
\omega(<\Omega, T^{A_{1},\dots,A_{n}}(X)\Omega>) \leq
\sum_{l=1}^{n} \omega(A_{l}) - 4(n-1)
\label{power}
\ee
where by
$\omega(d)$
we mean the order of singularity of the (numerical) distribution $d$ and by
$\omega(A)$
we mean the canonical dimension of the Wick monomial $W$.

\item
{\bf Wick expansion property}: In analogy to (\ref{comm-wick}) we require
\bea
[ \xi_{p}(y), T(A_{1}(x_{1}),\dots,A_{n}(x_{n})) ]
\nonumber\\
= - i~\sum_{m=1}^{n}~\prod_{l \leq m} (-1)^{|\xi_{p}||A_{l}|}~\sum_{q}~D_{pq}(y - x_{m} )~
T(A_{1}(x_{1}),\dots,\xi_{q}\cdot A_{m}(x_{m}),\dots, A_{n}(x_{n}))
\nonumber\\
\label{wick}
\eea
\end{itemize}

Up to now, we have defined the chronological products only for self-adjoint Wick monomials 
$
W_{1},\dots,W_{n}
$
but we can extend the definition for Wick polynomials by linearity.

The construction of Epstein-Glaser is based on the following recursive procedure. Suppose that
we know the chronological products up to order $n - 1$. Then we define the following expression:
\be
D(N) \equiv - \sum_{(X,Y) \in Part(N)}~( - 1)^{|Y|}~\epsilon(X,Y)~[ \bar{T}(X), T(Y) ]
\label{D-n}
\ee
where the partitions
$
(X,Y)
$
are restricted by
$
n \in X, Y \not= \emptyset,
$
$
|Y|
$
is the cardinal of $Y$ and the commutator is graded. These restrictions guarantee that
$
|X|, |Y| < n
$
so the expressions in the right-hand side of the previous expression are known by the induction
hypothesis. It is usual to denote
\bea
A^{\prime}(N) \equiv \sum_{(X,Y) \in Part(N)}~( - 1)^{|X|}~\epsilon(X,Y)~T(Y)~\bar{T}(X)
\nonumber\\
R^{\prime}(N) \equiv \sum_{(X,Y) \in Part(N)}~( - 1)^{|X|}~\epsilon(X,Y)~\bar{T}(X)~T(Y)
\eea
so 
\be
D(N) = A^{\prime}(N) - R^{\prime}(N). 
\ee
Then it can be proved that the expression
$
D(N) = D(A_{1}(x_{1}),\dots,A_{n}(x_{n}))
$
has causal support in the variables
$
x_{1} - x_{n},\dots,x_{n-1} - x_{n}
$
; accordingly is called the {\it causal commutator}. One can causally split 
$
D(N)
$
as
\be
D(N) = D^{\rm adv}(N) - D^{\rm ret}(N)
\ee
with $D^{\rm adv}(N)$ (resp. $D^{\rm ret}(N)$) with support in the upper (resp. lower) light cone and preserving power counting.
From these expression one can construct the chronological products
$
T(N)
$
in order $n$ in a standard way:
\be
T(N) = D^{\rm adv}(N) - A^{\prime}(N) = D^{\rm adv}(N) -  \sum_{(X,Y) \in Part(N)}~( - 1)^{|X|}~\epsilon(X,Y)~T(Y)~\bar{T}(X)
\ee

We can derive from these axioms the following result \cite{Sto1}, \cite{ward}, \cite{Bo}.
\begin{thm}
One can fix the causal products such that the following formula is true
\bea
T(\xi_{p}(y), A_{1}(x_{1}),\dots,A_{n}(x_{n}))
\nonumber\\
= - i~\sum_{m=1}^{n}~\prod_{l \leq m} (-1)^{|\xi_{p}||A_{l}|}~\sum_{q}~D^{F}_{pq}(y - x_{m} )~
T(A_{1}(x_{1}),\dots,\xi_{q}\cdot A_{m}(x_{m}),\dots, A_{n}(x_{n}))
\nonumber\\
+ \xi^{(+)}_{p}(y)~T(A_{1}(x_{1}),\dots, A_{n}(x_{n}))
+ \prod_{l \leq n}~(-1)^{|\xi_{p}| f_{l}}~T(A_{1}(x_{1}),\dots, A_{n}(x_{n}))~\xi^{(-)}_{p}(y)
\label{linear}
\eea
where 
$
D^{F}_{pq}
$
is a Feynman propagator associated to the causal distribution
$
D_{pq}
$.
\end{thm}
{\bf Proof:} Is done by induction on $n$. For 
$
n = 1, 2
$
it follows by direct computation. We suppose that it is valid for 
$
n = 1,\dots, N-1
$
and we go to
$
k = N.
$
Let us consider both sides of (\ref{linear}) for 
$
(y,x_{1},\dots,x_{N})
$
outside the main diagonal
$
D_{N+1}.
$
Using Bogoliubov axioms we prove that (\ref{linear}) is true in this domain. So, in general (\ref{linear}) 
can be broken by an anomaly
$
{\cal A}(y,x_{1},\dots,x_{N})
$
with support in
$
D_{N+1}
$
and appropriate symmetry properties. This anomaly can be eliminated by a redefinition of
$
T(\xi_{p}(y), A_{1}(x_{1}),\dots,A_{N}(x_{N})).
$
$\qed$
\newpage
\section{Yang-Mills Fields\label{ym}}

We consider a vector space 
$
{\cal H}
$
of Fock type generated (in the sense of Borchers theorem) by the vector field 
$
v_{\mu}
$ 
(with Bose statistics) and the scalar fields 
$
u, \tilde{u}
$
(with Fermi statistics). The Fermi fields are usually called {\it ghost fields}.
We suppose that all these (quantum) fields are of null mass. Let $\Omega$ be the
vacuum state in
$
{\cal H}.
$
In this vector space we can define a sesquilinear form 
$<\cdot,\cdot>$
in the following way: the (non-zero) $2$-point functions are by definition:
\bea
<\Omega, v_{\mu}(x_{1}) v_{\nu}(x_{2})\Omega> =i~\eta_{\mu\nu}~D_{0}^{(+)}(x_{1}
- x_{2}),
\nonumber \\
<\Omega, u(x_{1}) \tilde{u}(x_{2})\Omega> =- i~D_{0}^{(+)}(x_{1} - x_{2})
\nonumber\\
<\Omega, \tilde{u}(x_{1}) u(x_{2})\Omega> = i~D_{0}^{(+)}(x_{1} - x_{2})
\eea
and the $n$-point functions are generated according to Wick theorem. Here
$
\eta_{\mu\nu}
$
is the Minkowski metrics (with diagonal $1, -1, -1, -1$) and 
$
D_{0}^{(+)}
$
is the positive frequency part of the Pauli-Jordan distribution
$
D_{0}
$
of null mass. To extend the sesquilinear form to
$
{\cal H}
$
we define the conjugation by
\be
v_{\mu}^{\dagger} = v_{\mu}, \qquad 
u^{\dagger} = u, \qquad
\tilde{u}^{\dagger} = - \tilde{u}.
\ee

Now we can define in 
$
{\cal H}
$
the operator $Q$ according to the following formulas:
\bea
~[Q, v_{\mu}] = i~\partial_{\mu}u,\qquad
[Q, u] = 0,\qquad
[Q, \tilde{u}] = - i~\partial_{\mu}v^{\mu}
\nonumber \\
Q\Omega = 0
\label{Q-0}
\eea
where by 
$
[\cdot,\cdot]
$
we mean the graded commutator. One can prove that $Q$ is well defined. Indeed, we have the causal commutation relations 
\be
~[v_{\mu}(x_{1}), v_{\mu}(x_{2}) ] =i~\eta_{\mu\nu}~D_{0}(x_{1} - x_{2})~\cdot {\bf 1},
\qquad
[u(x_{1}), \tilde{u}(x_{2})] = - i~D_{0}(x_{1} - x_{2})~\cdot {\bf 1}
\ee
and the other commutators are null. The operator $Q$ should leave invariant
these relations, in particular 
\be
[Q, [ v_{\mu}(x_{1}),\tilde{u}(x_{2})]] + {\rm cyclic~permutations} = 0
\ee
which is true according to (\ref{Q-0}). The usefulness of
this construction follows from:
\begin{thm}
The operator $Q$ verifies
$
Q^{2} = 0.
$ 
The factor space
$
Ker(Q)/Ran(Q)
$
is isomorphic to the Fock space of particles of zero mass and helicity $1$
(photons). 
\end{thm}

$Q$ is the {\it gauge charge} for the multi-photon system.
The situation described above are susceptible of the following generalization. We can consider a system of 
$r$ species of particles of null mass and helicity $1$ if we use $r$ triplets
$
(v^{\mu}_{a}, u_{a}, \tilde{u}_{a}), a \in I
$
of massless fields; here $I$ is a set of indexes of cardinal $r$.
All the relations above have to be modified by appending an index $a$ to all these fields. 
\bea
~[Q, v^{\mu}_{a}] = i~\partial^{\mu}u_{a},\qquad
[Q, u_{a}] = 0,\qquad
~[Q, \tilde{u}_{a}] = - i~\partial_{\mu}v^{\mu}_{a}
\nonumber \\
Q\Omega = 0.
\label{Q-general}
\eea
We call this a pure Yang-Mills theory. This situation corresponds to QCD-type theories if the index $a$ carries a representation
of the group
$
SU(r).
$
Next we a looking for an interaction Lagrangian in the sense of (\ref{gauge-T}), tri-linear in the fields, of canonical dimension
$
\omega(T) \leq 4
$
and ghost number
$
gh(T) = 0.
$
One can prove that:

(i) $T$ is (relatively) cohomologous to a non-trivial co-cycle of the form:
\be
T = f_{abc} \left( \frac{1}{2}~:v_{a\mu}~v_{b\nu}~F_{c}^{\nu\mu}:
+ :u_{a}~v_{b}^{\mu}~\partial_{\mu}\tilde{u}_{c}:\right)
\label{T-sm}
\ee

(ii) The relation 
$
d_{Q}T = i~\partial_{\mu}T^{\mu}
$
is verified by:
\be
T^{\mu} = f_{abc} \left( :u_{a}~v_{b\nu}~F^{\nu\mu}_{c}: -
\frac{1}{2} :u_{a}~u_{b}~\partial^{\mu}\tilde{u}_{c}: \right)
\label{Tmu}
\ee

(iii) The relation 
$
d_{Q}T^{\mu} = i~\partial_{\nu}T^{\mu\nu}
$
is verified by:
\be
T^{\mu\nu} \equiv \frac{1}{2} f_{abc}~:u_{a}~u_{b}~F_{c}^{\mu\nu}:
\ee
where
\be
F_{a}^{\mu\nu} \equiv \partial^{\mu}v_{a}^{\nu} - (\mu \leftrightarrow \nu).
\label{F}
\ee
 
It is convenient to use the compact notation
$
T^{I}
$
where $I$ is a collection of indexes
$
I = [\nu_{1},\dots,\nu_{p}]~(p = 0,1,\dots,)
$
and the brackets emphasize the complete antisymmetry in these indexes. All these
polynomials have the same canonical dimension
\be
\omega(T^{I}) = \omega_{0},~\forall I
\ee
and because the ghost number of
$
T \equiv T^{\emptyset}
$
is supposed null, then we also have:
\be
gh(T^{I}) = |I|.
\ee
One can write compactly the previous relations as follows:
\be
d_{Q}T^{I} = i~\partial_{\mu}T^{I\mu}.
\label{gauge-1}
\ee

To be able to use the framework from the preceding two Sections we must find the corresponding algebra
$
\Xi
$
and generalize the previous construction for it. We take as variables 
$
\xi_{a}
$
the set of Grassmann variables
$
(v^{\mu}_{a}, u_{a}, \tilde{u}_{a}), a \in I
$
where 
$
v^{\mu}_{a}
$
are even and
$
u_{a}, \tilde{u}_{a}
$
are odd. The first-order jet extensions are 
$
(v^{\mu}_{a;\nu}, u_{a;\nu}, \tilde{u}_{a;\nu})
$
and (\ref{F}) can be easily defined in this context. The polynomial expressions
$
T^{I}
$
depend only on the variables
$
u_{a}, v_{a}^{\mu}
$
and the jet extensions
$
u_{a;\nu}, F_{a}^{\mu\nu}
$
as in the previous formulas where, instead of quantum fields we consider jet variables. 
There is an obvious way to write (\ref{Q-general}): we have a derivative operator
$
d_{Q}: \Xi \rightarrow \Xi
$
defined according to
\be
d_{Q}v^{\mu}_{a} = i~\eta^{\mu\nu}~u_{a;\nu},\qquad
d_{Q}u_{a} = 0,\qquad
d_{Q}\tilde{u}_{a} = - i~v^{\mu}_{a;\mu}.
\ee
This operator does not square to zero. Nevertheless we define
\be
\delta T^{I} \equiv \partial_{\mu}T^{I\mu} 
\ee
with
$
\partial_{\mu}
$
the formal derivative (see the beginning of Section \ref{wick prod}) and then
\be
s \equiv d_{Q} - i~\delta.
\ee

Now we can construct the chronological products
$
T(T^{I_{1}}(x_{1}),\dots,T^{I_{n}}(x_{n}))
$
according to the recursive procedure. Here the entries
$
T^{I_{1}},\dots,T^{I_{n}}
$
are considered as polynomials on 
$
\Xi.
$
We say that the theory is gauge invariant
in all orders of the perturbation theory if the following set of identities
generalizing (\ref{gauge-1}):
\be
d_{Q}T(T^{I_{1}}(x_{1}),\dots,T^{I_{n}}(x_{n})) = 
i \sum_{l=1}^{n} (-1)^{s_{l}} {\partial\over \partial x^{\mu}_{l}}
T(T^{I_{1}}(x_{1}),\dots,T^{I_{l}\mu}(x_{l}),\dots,T^{I_{n}}(x_{n}))
\label{gauge-n}
\ee
are true for all 
$n \in \N$
and all
$
I_{1}, \dots, I_{n}.
$
Here we have defined
\be
s_{l} \equiv \sum_{j=1}^{l-1} |I|_{j}.
\ee
In particular, the case
$
I_{1} = \dots = I_{n} = \emptyset
$
it is sufficient for the gauge invariance of the scattering matrix, at least
in the adiabatic limit: we have the same argument as for relation (\ref{phys-inv}).

To describe this property in a cohomological framework, we consider that the chronological products are the 
cochains and we define for the operator $\delta$ by 
\be
\delta T(T^{I_{1}}(x_{1}),\dots,T^{I_{n}}(x_{n})) = 
i \sum_{l=1}^{n} (-1)^{s_{l}} {\partial\over \partial x^{\mu}_{l}}
T(T^{I_{1}}(x_{1}),\dots,T^{I_{l}\mu}(x_{l}),\dots,T^{I_{n}}(x_{n})).
\label{der}
\ee
It is easy to prove that we have:
\be
\delta^{2} = 0
\ee
and
\be
[ d_{Q}, \delta ] = 0.
\ee
Next we define 
\be
s \equiv d_{Q} - i \delta
\ee
such that relation (\ref{gauge-n}) can be rewritten as
\be
sT(T^{I_{1}}(x_{1}),\dots,T^{I_{n}}(x_{n})) = 0.
\label{s-gauge-n}
\ee

We note that if we define
\be
\bar{s} \equiv d_{Q} + i \delta
\ee
we have
\be
s\bar{s} = 0, \qquad \bar{s} s = 0
\label{s2}
\ee
so expressions verifying the relation
$
s C = 0
$
can be called {\it cocycles} and expressions of the type
$
\bar{s}B
$
are the {\it coboundaries}. One can build the corresponding cohomology space in the standard way.

If we have (\ref{s-gauge-n}) for 
$
n = 1,2,\dots, n_{0} - 1
$
then the relation (\ref{s-gauge-n}) for
$
n_{0}
$
can be broken by anomalies i.e.we have:
\be
sT(T^{I_{1}}(x_{1}),\dots,T^{I_{n}}(x_{n})) = {\cal A}^{I_{1},\dots,I_{n}}(x_{1},\dots,x_{n})
\label{s-gauge-n-ano}
\ee
for 
$
n = n_{0};
$
here 
$
{\cal A}^{I_{1},\dots,I_{n}}
$
is a quasi-local expression, having support in
\be
D_{n} = \{ x_{1} = x_{2} = \dots = x_{n} \}.
\ee

The gauge theory is physically meaningful if one can remove the anomalies by a redefinition of the chronological products.
We will try to generalize (\ref{s-gauge-n}) admitting entries submonomials of
$
T^{I}
$
also.

\newpage

\section{Wick Submonomials and Gauge Invariance}{\label{ws}}
We recall that for a pure Yang-Mills theory the relevant Grassmann variables are:
$$
\xi = u_{a}, v^{\rho}_{a}, F^{\rho\sigma}_{a}, \tilde{u}_{a;\rho}
$$
(see the preceding Section). Then the derivative (\ref{derivative}) are in this case:
\bea
u_{a}\cdot T = f_{abc}~v_{b}^{\mu}~\partial_{\mu}\tilde{u}_{c}
\nonumber\\
u_{a}\cdot T^{\mu} = - f_{abc}~( v_{b\nu}~F^{\nu\mu}_{c} - u_{b}~\partial^{\mu}\tilde{u}_{c} )
\nonumber\\
u_{a}\cdot T^{\mu\nu} = f_{abc}~u_{b}~F^{\mu\nu}_{c}
\eea
\bea
v^{\rho}_{a}\cdot T = f_{abc}~(v_{b\sigma}F^{\sigma\rho}_{c} - u_{b}~\partial^{\rho}\tilde{u}_{c} )
\nonumber\\
v^{\rho}_{a}\cdot T^{\mu} = f_{abc}~u_{b}~F^{\mu\rho}_{c} 
\nonumber\\
v^{\rho}_{a}\cdot T^{\mu\nu} = 0
\eea
\bea
F^{\rho\sigma}_{a}\cdot T = - f_{abc}~v_{b}^{\rho}~v^{\sigma}_{c}
\nonumber\\
F^{\rho\sigma}_{a}\cdot T^{\mu} = f_{abc}~( \eta^{\mu\sigma}~u_{b}~v^{\rho}_{c} - \eta^{\mu\rho}~u_{b}~v^{\sigma}_{c} )
\nonumber\\
F^{\rho\sigma}_{a}\cdot T^{\mu\nu} = {1 \over 2}~f_{abc}~
(\eta^{\mu\rho} \eta^{\nu\sigma} - \eta^{\nu\rho}\eta^{\mu\sigma} )~u_{b}~u_{c}
\eea
\bea
\tilde{u}_{a;\rho}\cdot T = - f_{abc}~u_{b}~v_{c\rho}
\nonumber\\
\tilde{u}_{a;\rho}\cdot T^{\mu} = {1 \over 2}~f_{abc}~\delta^{\mu}_{\rho}~u_{b}~u_{c}
\nonumber\\
\tilde{u}_{a;\rho}\cdot T^{\mu\nu} = 0
\eea

The extra-sign from (\ref{derivative}) can be justified by the following off-shell relations:
\bea
s u_{a} \cdot T = - i~f_{abc}~(v_{b}^{\mu}~Kv_{c\mu} + u_{b}~K\tilde{u}_{c})
\nonumber\\
s u_{a} \cdot T^{\mu} = i~f_{abc}~u_{b}~Kv_{c}^{\mu}
\eea
\bea
s v_{a}^{\mu} \cdot T = - i~f_{abc}~u_{b}~Kv_{c}^{\mu}
\eea
and all other expressions
$
s \xi \cdot T^{I} 
$
null. We also have
\be
s F_{a}^{\rho\sigma} \cdot T^{I} = i~(-1)^{|I|}~ u_{a} \cdot T^{I\mu\nu}.
\ee

Here $K$ is the formal d'Alembert operator
\be
K \equiv \partial_{\mu}\partial^{\mu}
\ee
with
$
\partial_{\mu}
$
formal derivatives.

We now consider the gauge invariance in the second order of the perturbation theory for chronological
products having a derivative factor. By direct and long computation we can establish
\begin{thm}
The  chronological products from the second order of the perturbation theory can be chosen such that the tree contributions
verify:
\bea
sT(u_{a}\cdot T^{I}(x),T^{J}(y)) = 0
\nonumber\\
sT(F_{a}^{\rho\sigma}\cdot T^{I}(x),T^{J}(y)) = i~T(u_{a}\cdot T^{I\mu\nu}(x),T^{J}(y))
\nonumber\\
sT(\tilde{u}_{a}^{\mu}\cdot T^{I}(x),T^{J}(y)) = 0
\eea
if we perform the following finite renormalizations:
\bea
N(u_{a}\cdot T^{\mu}(x),T(y)) = i~\delta(x - y)~f_{abc}~f_{dec}~(v_{b\nu}~v_{d}^{\nu}~v_{e}^{\mu})(x)
\nonumber\\
N(u_{a}\cdot T^{\mu}(x),T^{\nu}(y)) = i~\delta(x - y)~f_{abc}~f_{dec}~(v_{b}^{\nu}~u_{d}~v_{e}^{\mu})(x)
\nonumber\\
N(u_{a}\cdot T^{\mu\nu}(x),T(y)) = i~\delta(x - y)~f_{abc}~f_{dec}~(u_{b}~v_{d}^{\mu}~v_{e}^{\nu})(x)
\nonumber\\
N(u_{a}\cdot T^{\mu\nu}(x),T^{\rho}(y)) = - i~\delta(x - y)~f_{abc}~f_{dec}~
[ \eta^{\mu\rho}~u_{b}~u_{d}~v_{e}^{\nu} - (\mu \leftrightarrow \nu)] (x)
\nonumber\\
N(u_{a}\cdot T^{\mu}(x),T^{\rho\sigma}(y)) = \frac{i}{2}~\delta(x - y)~f_{abc}~f_{dec}~
[\eta^{\mu\rho}~v_{b}^{\sigma}~u_{d}~u_{e} - (\rho \leftrightarrow \sigma)] (x)
\eea
\end{thm}

The proof is done using the off-shell method from \cite{off}. First one computes the off-shell tree contributions 
of the commutators
$
D(\xi\cdot T^{I}(x),T^{J}(y)) = [ \xi\cdot T^{I}(x),T^{J}(y)]
$
and obtains expressions of the type
$
D(x - y), \partial D(x - y),
$
etc. multiplying Wick polynomials. Next we compute
$
sD(\xi\cdot T^{I}(x),T^{J}(y))
$
and only terms involving distributions of the type
$
KD(x - y), \partial KD(x - y),
$
etc. survive. If we perform the causal splitting (see the previous Section) making the substitutions
$
D \rightarrow D^{adv,ret}, \partial D(x - y) \rightarrow \partial D^{adv,ret},
$
etc. then we obtain anomalies because on-shell we have
$
K~D(x - y) = 0
$
but
$
K~D^{adv,ret}(x - y) = \delta(x - y).
$
These anomalies can be eliminated performing the finite renormalizations described above. 
The loop contributions in the previous relations can be investigated as in \cite{sr2} and they do not produce anomalies.

We can write in a compact way the content of the this theorem as follows
\be
s T(\xi\cdot T^{I}(x), T^{J}(y)) = T(s\xi\cdot T^{I}(x), T^{J}(y)),\quad \xi = u_{a}, F^{\rho\sigma}_{a}, \tilde{u}_{a;\rho}.
\label{G(1,2)}
\ee

Now we go to chronological products with two derivative entries. We have in the same way:
\begin{thm}
The  chronological products from the second order of the perturbation theory can be chosen such that the tree contributions
verify:
\bea
sT(u_{a}\cdot T^{I}(x),u_{b}\cdot T^{J}(y)) = - (-1)^{|I|(|J|+1)}~\delta(x - y)~f_{abc}~u_{c}\cdot T^{IJ}(x)
\nonumber\\
\nonumber\\
sT(u_{a}\cdot T^{I}(x),F_{b}^{\mu\nu}\cdot T^{J}(y)) = 
\nonumber\\
- (-1)^{|I|(|J|+1)}~\delta(x - y)~f_{abc}~F_{c}^{\mu\nu}\cdot T^{IJ}(x)
- (-1)^{|J|}~T(u_{a}\cdot T^{I}(x),u_{b}\cdot T^{J\mu\nu}(y))
\nonumber\\
\nonumber\\
sT(u_{a}\cdot T^{I}(x),\tilde{u}_{b;\rho}\cdot T^{J}(y)) = - (-1)^{|I|(|J|+1)}~\delta(x - y)~f_{abc}~\tilde{u}_{c;\rho}\cdot T^{IJ}(x)
\nonumber\\
\nonumber\\
sT(F_{a}^{\mu\nu}\cdot T^{I}(x),F_{b}^{\rho\sigma}\cdot T^{J}(y)) = 
\nonumber\\
i~(-1)^{|I|}~[ T(u_{a}\cdot T^{I\mu\nu}(x),F_{b}^{\rho\sigma}\cdot T^{J}(y))
+ (-1)^{|J|}~T(F_{a}^{\mu\nu}\cdot T^{I}(x),u_{b}\cdot T^{J\rho\sigma}(y))
\nonumber\\
\nonumber\\
sT(F_{a}^{\mu\nu}\cdot T^{I}(x),\tilde{u}_{b;\rho}\cdot T^{J}(y)) = 
i~(-1)^{|I|}~T(u_{a}\cdot T^{I\mu\nu}(x),\tilde{u}_{b;\rho}\cdot T^{J}(y))
\nonumber\\
\nonumber\\
sT(\tilde{u}_{a;\rho}\cdot T^{I}(x),\tilde{u}_{b;\sigma}\cdot T^{J}(y)) = 0
\eea
if we perform the following finite renormalizations:
\bea
N(u_{a}\cdot T^{\mu}(x),u_{b}\cdot T^{\nu}(y)) = 
i~\delta(x - y)~f_{adc}~f_{bec}~(- v_{d}^{\nu}~v_{e}^{\mu} + \eta^{\mu\nu}~v_{d\rho}~v_{e}^{\rho})(x)
\nonumber\\
N(u_{a}\cdot T^{\mu\nu}(x),u_{b}\cdot T^{\rho}(y)) = 
i~\delta(x - y)~f_{adc}~f_{bec}~[\eta^{\mu\rho}~u_{d}~v_{e}^{\nu} - (\mu \leftrightarrow \nu)](x)
\nonumber\\
N(u_{a}\cdot T^{\mu\nu}(x),u_{b}\cdot T^{\rho\sigma}(y)) = i~\delta(x - y)~f_{abc}~f_{dec}~
[\eta^{\mu\rho}~\eta^{\nu\sigma} - (\mu \leftrightarrow \nu)](u_{d}~u_{e})(x)
\eea
\end{thm}

The loop contributions in the previous relations can be investigated as in \cite{sr2} and they do not produce anomalies.
We can write in a compact way the content of the previous theorem if we introduce the following product 
$
\circ : \Xi \times \Xi \rightarrow \Xi
$ 
according to
\bea
u_{a} \circ u_{b} = f_{abc}~u_{c}
\nonumber \\
u_{a} \circ F^{\rho\sigma}_{b} = f_{abc}~F^{\rho\sigma}_{c}
\nonumber \\
u_{a} \circ \tilde{u}_{b;\rho} = f_{abc}~\tilde{u}_{c;\rho}
\eea
We also have Grassmann commutativity
\be
\xi \circ \eta = (-1)^{|\xi||\eta|} \eta \circ \xi
\ee
and the rest of the products are null. Then we have
\bea
s T(\xi\cdot T^{I}(x), \eta\cdot T^{J}(y)) = 
T(s\xi\cdot T^{I}(x), \eta\cdot T^{J}(y)) + (-1)^{|I| + |\xi|}~T(\xi\cdot T^{I}(x), s\eta\cdot T^{J}(y))
\nonumber\\
+ \epsilon^{IJ}(\xi,\eta)~\delta(x - y)~\xi \circ \eta \cdot T^{IJ}(x),
\quad \xi, \eta = u_{a}, F^{\rho\sigma}_{a}, \tilde{u}_{a;\rho}.
\label{G(2,2)}
\eea
where the sign is
\be
\epsilon^{IJ}(\xi,\eta) = (-1)^{s},\qquad
s = (|I| + |\xi| + 1)~(|J| + 1) + |I|.
\label{epsi}
\ee

We want to extend (\ref{G(1,2)}) and (\ref{G(2,2)}) for the case 
$
\xi, \eta = v^{\mu}_{a}.
$
To do this we must enlarge the algebra $\Xi$ to the algebra
$
\Xi_{\rm ext}
$
adding the variables
$
Y_{a}
$
and
$
R^{\mu\nu}_{a}
$
defined according to their action on elements $A$ from the algebra $\Xi$:
\bea
Y_{a} \cdot A^{I} = {1\over 4}~(-1)^{|A| + |I|}~~\tilde{u}_{a;\alpha}\cdot A^{I\alpha}
\nonumber\\
R^{\mu\nu}_{a} \cdot A^{I} = u_{a} \cdot A^{I\mu\nu}.
\eea

Then we must postulate new non-zero $\circ$ products
\bea
u_{a} \circ Y_{b} = f_{abc}~Y_{c}
\nonumber\\
u_{a} \circ R^{\mu\nu}_{b} = f_{abc}~R^{\mu\nu}_{c}
\nonumber\\
F^{\rho\sigma}_{a} \circ R^{\mu\nu}_{b} = f_{abc}~(\eta^{\mu\rho}~\eta^{\nu\sigma} - \eta^{\nu\rho}~\eta^{\mu\sigma} ) Y_{c}.
\eea
and
\bea
v^{\mu}_{a} \circ  u_{b} = - f_{abc}~v_{c}^{\mu}
\nonumber\\
v^{\mu}_{a} \circ  F_{b\rho\sigma} = 
f_{abc}~[\delta^{\mu}_{\rho}~\tilde{u}_{c;\sigma} - (\rho \leftrightarrow \sigma) ])
\nonumber\\
v^{\mu}_{a} \circ \tilde{u}_{b;\nu} = \delta^{\mu}_{\nu}~f_{abc}~Y_{c}
\eea
\be
v^{\mu}_{a} \circ v^{\nu}_{b}  = - f_{abc}~R^{\mu\nu}_{c}
\ee
together with Grassmann commutativity. Using these rules (\ref{G(1,2)}) and (\ref{G(2,2)}) are true for the case 
$
\xi, \eta = v^{\mu}_{a}
$
also.

The two products in 
$
\Xi_{\rm ext}
$
are connected by
\begin{lemma}
The following relation is true for all
$
\xi_{1}, \xi_{2}, \xi_{3} \in \Xi_{\rm ext}:
$
\be
(\xi_{1} \circ \xi_{2}) \cdot \xi_{3} + (-1)^{t_{1}} (\xi_{2} \circ \xi_{3}) \cdot \xi_{1} + (-1)^{t_{2}} (\xi_{3} \circ \xi_{1}) \cdot \xi_{2} = 0
\ee
where
\be
t_{1} = |\xi_{1}| + |\xi_{3}|~( |\xi_{1}| + |\xi_{2}| + 1 ), \qquad
t_{2} = |\xi_{3}| + |\xi_{2}|~( |\xi_{1}| + |\xi_{3}| + 1 ).
\ee
\end{lemma}

Moreover 
$
\Xi_{\rm ext}
$
becomes a graded Lie algebra with respect to the product $\circ$ because we have:
\begin{lemma}
The following relation is true for all
$
\xi_{1}, \xi_{2}, \xi_{3} \in \Xi_{\rm ext}:
$
\be
(\xi_{1} \circ \xi_{2}) \circ \xi_{3} + (-1)^{s_{1}} (\xi_{2} \circ \xi_{3}) \circ \xi_{1}
+ (-1)^{s_{2}} (\xi_{3} \circ \xi_{1}) \circ \xi_{2} = 0
\ee
where
\be
s_{1} = |\xi_{1}|~( |\xi_{2}| + |\xi_{3}|), \qquad
s_{2} = |\xi_{3}|~( |\xi_{1}| + |\xi_{2}|).
\ee
\end{lemma}

\newpage
\section{Gauge Invariance of Higher Rank\label{rank}}

In this Section we propose a conjecture which generalize (\ref{G(1,2)}) and (\ref{G(2,2)}). Let 
$
A^{I_{1}}_{1}, A^{I_{2}}_{2}
$
be of the form 
$
T^{I}
$
or
$
\xi \cdot T^{I},~\xi \in \Xi.
$
We define the product 
$
A^{I_{1}}_{1} \circ A^{I_{2}}_{2}
$
according to 
\bea
\xi_{1} \cdot T^{I_{1}} \circ \xi_{2} \cdot T^{I_{2}} = 
\epsilon^{I_{1}I_{2}}(\xi_{1},\xi_{2})~\xi_{1} \circ \xi_{2} \cdot T^{I_{1}I_{2}}
\nonumber\\
T^{I_{1}} \circ \xi_{2} \cdot T^{I_{2}} = \xi_{1} \cdot T^{I_{1}} \circ \cdot T^{I_{2}} = T^{I_{1}} \circ T^{I_{2}} = 0
\eea
where
$
\epsilon^{I_{1}I_{2}}(\xi_{1},\xi_{2})
$
has been defined in (\ref{epsi}). Then we have:
\begin{thm}
The Wick submonomials are a graded Lie algebra with respect to the product $\circ$:
\be
(A_{1} \circ A_{2}) \circ A_{3} + (-1)^{u_{1}} (A_{2} \circ A_{3}) \circ A_{1}
+ (-1)^{u_{2}} (A_{3} \circ A_{1}) \circ A_{2} = 0
\ee
where
\be
u_{1} = |A_{1}|~( |A_{2}| + |A_{3}|), \qquad
u_{2} = |A_{3}|~( |A_{1}| + |A_{2}|).
\ee
\end{thm}

We call {\it gauge invariance identities}:
\bea
sT(A^{I_{1}}_{1}(x_{1}),\dots,A^{I_{n}}_{n}(x_{n})) = \sum_{m=1}^{n}~\prod_{l < m}~(-1)^{f_{l}}~
T(A^{I_{1}}_{1}(x_{1}),\dots,sA^{I_{m}}_{m}(x_{m}),\dots,A^{I_{n}}_{n}(x_{n})) 
\nonumber\\
+ \sum_{1 \leq p < q \leq n}~\epsilon_{pq}(A^{I_{1}}_{1},\dots,A^{I_{n}}_{n})~\delta (x_{p} - x_{q} ) \times
\nonumber\\
T(A^{I_{p}}_{p} \circ A^{I_{q}}_{q}(x_{p}),A^{I_{1}}_{1}(x_{1}),\dots,\hat{p},\dots,\hat{q},\dots,A^{I_{n}}_{n}(x_{n}))
\label{G(r,n)}
\eea
where 
$
\epsilon_{pq}(A^{I_{1}}_{1},\dots,A^{I_{n}}_{n})
$
is the Fermi sign associated to the permutation
$$
(A_{1},\dots,A_{n}) \rightarrow (A_{p},A_{q},A_{1},\dots,\hat{p},\dots,\hat{q},\dots,A_{n}).
$$
Explicitly
\be
\epsilon_{pq}(A^{I_{1}}_{1},\dots,A^{I_{n}}_{n}) = \prod_{l=1, l \not= p}^{q-1}~(-1)^{f_{l}f_{q}}~
\prod_{l=1}^{p-1}~(-1)^{f_{l} f_{p}}
\ee
and
$
f_{l} = |A_{l}| + |I_{l}|.
$
If $r$ of the expressions
$
A_{1},\dots,A_{n}
$
are of the type
$
\xi \cdot T^{I}
$
then we say that the gauge identities are {\it of rank r} and we denote (\ref{G(r,n)}) by
$
G(r,n).
$
We can see that (\ref{G(1,2)}) corresponds to 
$
r =1, n = 2
$
and (\ref{G(2,2)}) corresponds to 
$
r = 2, n = 2.
$

It is clear that the case
$
r = 0
$
corresponds to the gauge invariance as postulated before i.e. (\ref{s-gauge-n}) is 
$
G(0,n).
$
We close by showing how one can use gauge invariance of higher rank to prove (\ref{s-gauge-n}).
We give the connection between 
$
G(1,n)
$
and (\ref{s-gauge-n}).
\begin{thm}
The following formula is true:
\bea
[ \xi_{p}(y), sT(T^{I_{1}}_{1}(x_{1}),\dots,T^{I_{n}}_{n}(x_{n})) ]
\nonumber\\
= - i~(-1)^{|\xi_{p}|}~\sum_{m=1}^{n}~\sum_{q}~D_{pq}(y - x_{m} )~\prod_{l < m} (-1)^{|I_{l}|||I_{m}|}~(-1)^{|\xi_{p}| |I_{m}|}~\times
\nonumber\\
\times [ sT(\xi_{q}\cdot T^{I_{m}}_{m}(x_{m}),T^{I_{1}}_{1}(x_{1}),\dots,\hat{m},\dots, T^{I_{n}}_{n}(x_{n}))
\nonumber\\
- T(s\xi_{q}\cdot T^{I_{m}}_{m}(x_{m}),T^{I_{1}}_{1}(x_{1}),\dots,\hat{m},\dots, T^{I_{n}}_{n}(x_{n})) ]
\eea
\end{thm}

From this formula it follows that if we have gauge invariance (\ref{s-gauge-n}) then we have gauge invariance of the first rank
$
G(1,n);
$
conversely, if we have 
$
G(1,n)
$
the the right hand side of the preceding formula  being zero we have
\be
[ \xi_{p}(y), sT(T^{I_{1}}_{1}(x_{1}),\dots,T^{I_{n}}_{n}(x_{n})) ] = 0
\ee
so
\be
sT(T^{I_{1}}_{1}(x_{1}),\dots,T^{I_{n}}_{n}(x_{n})) = {\cal A}^{I_{1},\dots,I_{n}}(x_{1},\dots,x_{n})
\ee
where the anomalies - see (\ref{s-gauge-n-ano}) - are numerical distribution (times ${\bf 1}$). 
So we are reduces to the study of pure numerical anomalies. In the same way one can pass from
$
G(2,n)
$
to
$
G(1,n)
$
and in general, from
$
G(r,n)
$
to
$
G(r - 1,n).
$
It is clear that this inductive process starts from 
$
G(n,n)
$
where we will have only tree and one-loop contributions. From power counting it follows that for  
$
n > 5
$ 
there are no anomalies in 
$
G(n,n).
$
To establish
$
G(n,n)
$
we still have to consider the cases
$
n = 3, 4, 5.
$
We have succeeded to prove the case
$
n = 3
$
using the techniques of \cite{loop} but one still has to consider the cases
$
n = 4, 5
$
where only one-loop contributions can give anomalies. 

\section{Conclusions}
One can generalize the previous framework to the full standard model including particles of spin $1$ and positive mass, Dirac fields
(describing matter) and scalar fields (of Higgs type) as described, for instance in \cite{loop}. It involves a careful generalization
of the product 
$\circ$.

It is not clear that the previous method will provide a proof of gauge invariance in all orders of the perturbation theory
$
G(0,n).
$
However, the method reduces this identity to ``simpler" ones
$
G(r,n),~r = n, n - 1, \dots, 1
$
and it is clear that even these ``simpler" identities are not easy to prove in general. We have tested cohomological methods
for the case 
$
n = 4
$
and one can eliminate the the anomalies from the even sector (with respect to parity) but not from the odd sector. So, it seems that one 
has to do explicit computations, as in \cite{loop}.

\end{document}